\documentclass{aa}

\usepackage{txfonts,makecell}
\usepackage{hhline}
\usepackage{hyperref}
\usepackage{cleveref}

\crefname{equation}{eq.}{eqs.}

\newcounter{mysfig}
\renewcommand\themysfig{(\alph{mysfig})}
\makeatletter
\newcommand\Scaption[1]{%
\refstepcounter{mysfig}%
\vskip.5\abovecaptionskip
  \sbox\@tempboxa{\small\themysfig~#1}%
  \ifdim \wd\@tempboxa >\hsize
    \small\themysfig~#1\par
  \else
    \global \@minipagefalse
    \hb@xt@\hsize{\hfil\box\@tempboxa\hfil}%
  \fi
  \vskip\belowcaptionskip}
\makeatother

\begin{document} 

  \title{Is there a concordance value for $H_0$?}
  
  \author{
    Vladimir V. Lukovi\'{c}\inst{\ref{inst1}}
    \and Rocco D'Agostino\inst{\ref{inst1}}
    \and Nicola Vittorio\inst{\ref{inst1},\ref{inst2}}
  }
  
  \institute{Dipartimento di Fisica, Universit\`{a} di Roma "Tor Vergata", Via della Ricerca Scientifica 1, I-00133, Roma, Italy\label{inst1}\and Sezione INFN, Universit\`{a} di Roma "Tor Vergata", Via della Ricerca Scientifica 1, I-00133, Roma, Italy\label{inst2}}

  \offprints{vladimir.lukovic@roma2.infn.it}

  \date{Received / Accepted }

  \abstract{We test the theoretical predictions of several cosmological models against different observables to compare the indirect estimates of the current expansion rate of the Universe determined from model fitting with the direct measurements based on Cepheids data published recently.}
  {We perform a statistical analysis of type Ia supernova (SN Ia), Hubble parameter, and baryon acoustic oscillation data. A joint analysis of these datasets allows us to better constrain cosmological parameters, but also to break the degeneracy that appears in the distance modulus definition between $H_0$ and the absolute B-band magnitude of SN Ia, $M_0$.}
  {From the theoretical side, we considered spatially flat and curvature-free $\Lambda$CDM, $w$CDM, and inhomogeneous Lema\^{i}tre-Tolman-Bondi (LTB) models. To analyse SN Ia we took into account the distributions of SN Ia intrinsic parameters.}
  {For the $\Lambda$CDM model we find that $\Omega_m=0.35\pm0.02$, $H_0=(67.8 \pm 1.0)\,$km$\,$s$^{-1}/$Mpc, while the corrected SN absolute magnitude has a normal distribution ${\cal N}(19.13,0.11)$. The $w$CDM model provides the same value for $\Omega_m$, while $H_0=(66.5\pm1.8)\,$km$\,$s$^{-1}/$Mpc and $w=-0.93\pm0.07$. When an inhomogeneous LTB model is considered, the combined fit provides $H_0=(64.2 \pm 1.9)\,$km$\,$s$^{-1}/$Mpc.}
  {Both the Akaike information criterion and the Bayes factor analysis cannot clearly distinguish between $\Lambda$CDM and $w$CDM cosmologies, while they clearly disfavour the LTB model. For the $\Lambda$CDM, our joint analysis of the SN Ia, the Hubble parameter, and the baryon acoustic oscillation datasets provides $H_0$ values that are consistent with cosmic microwave background
(CMB)-only Planck measurements, but they differ by $2.5\sigma$ from the value based on Cepheids data.}

  \keywords{cosmology: cosmological parameters, distance scale, dark matter, dark energy}
  
  \authorrunning{V. V. Lukovi\'{c}, R. D'Agostino, N. Vittorio}
 
  \maketitle
  
  \section{Introduction}
    \label{sec:int}
Since the early determination by Hubble \citep{Hubble29}, the Hubble constant was for a long time believed to be between 50 and 100 km$\,$s$^{-1}/$Mpc \citep{Kirschner03}. Recent findings are obtained by means of space facilities, improved control of systematics, and the use of different calibration techniques, as in the Hubble Space Telescope Key Project, which estimated $H_0=(72 \pm 8)\,$km$\,$s$^{-1}/$Mpc \citep{HST01}. \citet{Riess16} provided the most recent direct estimate of the expansion rate of the Universe: $H_0=(73.0 \pm 1.8)\,$km$\,$s$^{-1}/$Mpc. Together with these extraordinary improvements in the direct determination of the distance ladder, there are by now different classes of observations that allow an indirect estimate of the Hubble constant. Among others, the observations of the cosmic microwave background (CMB) anisotropy by WMAP \citep{WMAP9} and \citet{Planck15} satellites yielded values of $H_0=(70.0 \pm 2.2)$ km$\,$s$^{-1}/$Mpc and $H_0=(67.27 \pm 0.66)$ km$\,$s$^{-1}/$Mpc, respectively. In addition to the CMB anisotropy measurements, other observables have been crucial to constrain the cosmological parameters, such as type Ia supernovae (SN Ia). The high-z supernova search team led by Adam Riess together with Brian P. Schmidt \citep{Riess98} and the supernova cosmology project led by Saul Perlmutter \citep{Perlmutter99} reported the first evidence for an accelerated cosmic expansion. Since then, the number of observed SN Ia increased by about an order of magnitude. Different publicly available compilations have been used to constrain cosmological models: Union2 \citep{Amanullah10}, Union2.1 \citep{Suzuki12}, Constitution set \citep{Hicken09}, and JLA \citep{Betoule14}. The results confirm the need for a late accelerated expansion of the Universe, consistent with the findings of the WMAP  and Planck missions. Unfortunately, the observations of SN Ia by themselves are not able to provide a value for the local expansion rate of the Universe, $H_0$, since this parameter is degenerate with the SN absolute magnitude. However, there are other cosmological observables that are more directly sensitive to the value of the Hubble constant. On one hand, passively evolving red galaxies, which are dominated by the older stellar population, whose age can be accurately estimated from a spectroscopic analysis (also known as cosmic chronometers), can be used to provide the redshift dependence of the expansion rate, $H(z)$, as suggested by \citet{Jimenez02}. Fitting these observational Hubble data (OHD), \citet{Liu15} found a value of $H_0=67.6\,$km$\,$s$^{-1}/$Mpc. On the other hand, the baryon acoustic oscillation (BAO) data have been used to constrain the cosmological parameters, providing results that agree with the most recent findings of the Planck Collaboration. In particular, a recent estimate of the Hubble constant provides $H_0 = (68.11 \pm 0.86)$ km$\,$s$^{-1}/$Mpc \citep{Cheng15}.\\
\indent It is clear that the indirect estimates of the Hubble constant lead to lower values of $H_0$ compared to the direct measurements. Even the earlier estimate of $H_0=(73.8\pm2.4)\,$km$\,$s$^{-1}/$Mpc by \citet{Riess11} and the latest one \citep{Riess16} contradict the most recent result from Planck (TT, TE, EE + lowP) at  the $2.6\sigma$ and $3.0\sigma$ level, respectively. The question now is whether this difference hides new physics beyond what is by now commonly called the concordance model. This point has been addressed by \citet{Efstathiou14}, who reanalysed the Cepheid data used by \citet{Riess11}. He obtained a value $H_0=(72.5\pm2.5)$ km$\,$s$^{-1}/$Mpc, reducing the difference to Planck to only $2 \sigma$ and concluding that there is no evidence for new physics (see also \citet{Chen11} and \citet{Marra13}). We here extend this discussion to determine whether any difference is present when observables other than CMB are considered. To do so, we perform a separate and a joint analysis of SN Ia, OHD, and BAO data. The joint analysis promises to provide more stringent constraints on the cosmological models, and to break the degeneracy between the SN absolute magnitude and the Hubble constant, which is peculiar to the SN analysis. 

Several SN datasets (such as Union and Constitution) provide cosmological distance moduli that are derived assuming a flat $\Lambda$CDM model. Hence, these datasets need to be treated with caution when used to constrain cosmological models that are different from $\Lambda$CDM. We used the JLA dataset, which provides model-independent apparent magnitudes instead of model-dependent distance moduli. Moreover, the increase in the amount of data and the improvement in systematics imply that a more complete statistical analysis is necessary. We therefore followed the approach proposed by \citet{Nielsen15} for the SN data analysis. For the theoretical models we considered the standard flat $\Lambda$CDM model and its extensions, which include the curvature-free $k\Lambda$CDM model and a dark energy model characterised by an equation of state (EoS) $p=w\rho c^2$, with $w=const$. In addition, we also considered a different class of models, based on the Lema\^{i}tre-Tolman-Bondi (LTB) metric, which describes an isotropic but inhomogeneous Universe \citep{Lemaitre33,Tolman34,Bondi47,Krasinski97}, to stress the dependence of the Hubble constant estimates on the assumed theoretical model.

The plan of the paper is as follows. In \Cref{sec:th} we review the theoretical models we considered. In \Cref{sec:analysis} we review the observables and datasets used in our analysis. In \Cref{sec:res} we show the results of our comparison between theory and observations. Finally, in \Cref{sec:con} we  summarise our findings and conclusions.

  \section{Theoretical models}
    \label{sec:th}
    All the models considered here arise from the exact solutions of the Einstein field equations (EE) ${G^\mu}_\nu=\kappa {T^\mu}_\nu$, where ${G^\mu}_\nu$ is the Einstein tensor, $\kappa=8\pi G/c^4$ , and ${T^\mu}_\nu=\mathrm{diag}(\rho c^2, -p, -p, -p)$ is the form of the energy-momentum tensor for a perfect fluid in the comoving frame. Here $p$ and $\rho$  (pressure and density of the fluid) are related by the equation of state (EoS) $p=w\rho c^2$.
    \subsection{ Friedmann-Lema\^{i}tre-Robertson-Walker models}
      \noindent Friedmann-Lema\^{i}tre-Robertson-Walker (FLRW) models describe a homogeneous and isotropic Universe. Under such conditions, EE can be solved exactly, which results in the metric \citep{Friedmann22,Friedmann24,Robertson35,Walker37} 
      \begin{equation}
        ds^2=c^2 dt^2-R^2(t)\left[\dfrac{dr^2}{1-k r^2}+r^2(d\theta^2+\sin^2\theta\ d\phi^2)\right]
        \label{eq:1}
      ,\end{equation}
      where $R(t)$ is a scale factor in units of length and $k=-1,\ 0,\ +1$ is a curvature parameter for the open, flat, and closed 3D space geometry, respectively.
      The Hubble expansion rate as a function of redshift is defined as
      \begin{equation}
        H(z) \equiv d(\ln{R})/dt=-(1+z)^{-1} dz/dt
        \label{eq:defH}
      .\end{equation}
      Using the Friedmann equation, it can be expressed as 
      \begin{equation}
        H(z)=H_0 E(z)
        \label{eq:Hofz}
      ,\end{equation}
      where $H_0\equiv H(z=0)$, while the adimensional Hubble parameter $E(z)$ is given by
      \begin{equation}
        E(z)=\sqrt{\sum_i\Omega_i(1+z)^{3(1+w_i)}+\Omega_k(1+z)^2}
        \label{eq:E(z)}
      .\end{equation}
      Here $\Omega_k\equiv-k c^2/[H_0^2 R(t_0)^2]$ and $t_0$ is the age of the Universe, while the sum runs over all the components of the cosmological fluid, which are each characterised by its own EoS and density parameter, $\Omega_i\equiv\rho_i/\rho_c$, and the present density of the $i$-th component in units of the critical density $\rho_c=8\pi G/(3H_0^2)$. The functional dependence of the luminosity distance with the redshift is fixed by the cosmological model. In the FLRW model $d_L(z)$ is calculated according to the equation
      \begin{equation}
        d_L(z)=\dfrac{c(1+z)}{H_0\sqrt{|\Omega_k|}}S_k\left[\sqrt{|\Omega_k|}\int_0^z\dfrac{dz'}{E(z')}\right]
        \label{eq:FdL}
      ,\end{equation}
      where the function $S_k$ depends on the curvature, 
      \begin{equation}
        S_k(\tau)\equiv\left\{
        \begin{aligned}
          &\sin \tau &&\text{for}\,\,k=+1 \\
          &\tau  &&\text{for}\,\,k=0  \\
          &\sinh \tau  &&\text{for}\,\,k=-1
        \end{aligned}\right.
        \label{eq:21}
      .\end{equation}\Cref{eq:Hofz,eq:FdL} are used in \Cref{sec:res} to fit theoretical models to observables such as the Hubble expansion rate and the SN Ia.
        
        There is overwhelming evidence that about a quarter of the critical density in the Universe is in the form of a cold, weakly interacting dark matter (CDM) and that an extra component in the cosmological fluid is needed for closing the Universe. Although the physical nature of this dark energy (DE) component is poorly understood, it currently provides the only explanation for the accelerated expansion of the Universe in a FLRW cosmology \citep{Riess98, Perlmutter99}. The second Friedmann equation
      \begin{equation}
        \frac{\ddot a}a=-\frac{4\pi G}3\sum_i\rho_i(1+3w_i)
        \label{eq:Fd2}
      \end{equation}
       shows that for the cosmic fluid to be in an accelerated expansion, at least one component must have $w<-1/3$.
       The density evolution is provided by the time component of the conservation equations ${T^{\alpha\beta}}_{;\beta}=0$: 
      \begin{equation}
        \frac{d\rho}{d a} + \frac{3\rho} a \left(w +1\right) = 0
        \label{eq:rhovsw}
      .\end{equation}
       While DM is more constrained and commonly considered cold and pressureless ($w\equiv0$), the DE models consider various EoS for DE fluid. For $w\equiv-1$, the DE density is constant (cf. \Cref{eq:rhovsw}) and can be described in terms of a non-vanishing cosmological constant $\Lambda= 8\pi G\rho_\Lambda/c^2$. This case recovers the flat concordance $\Lambda$CDM model considered to be the simplest way of best-fitting current cosmological observations. We also consider a $w$CDM model with $-1<w<-1/3$, which we assume to be constant with respect to the cosmic expansion.
       
    \subsection{LTB models}
      To explain the accelerated expansion suggested by SN Ia observations, the homogeneous and isotropic FLRW model must resort to DE. However, the same effect can be explained in an alternative way, by relaxing the homogeneity requirement of the cosmological principle and presuming an isotropic Gpc-size underdensity in matter distribution (see e.g. \citet{Alnes06a}). Under pressureless conditions, the local isotropic, but inhomogeneous Universe is described by the LTB metric
      \begin{equation}
        ds^2=c^2 dt^2-\dfrac{{R'}^2(r,t)}{1-k(r)}dr^2-R^2(r,t)[d\theta^2+\sin^2\theta\ d\phi^2]
        \label{eq:9}
      ,\end{equation}
      where $k(r)$ determines the spatial curvature of 3D space. We denote derivatives with respect to the comoving radial coordinate $r$ and to the time $t$ with prime and dot, respectively. As before, $R(r,t)$ is the scale factor in units of length, and we can introduce the reduced scale factor as well: $a(r,t)\equiv R(r,t)/R_0(r)$, with $R_0(r)\equiv R(r,t_0)$. Unlike the standard model, the LTB model is characterised by radial and transverse expansion rates:
      \begin{align}
        H_\parallel(r,t)&\equiv\dfrac{\dot{R}'(r,t)}{R'(r,t)}
        \label{eq:Hpl}\\
        H_\perp(r,t)&\equiv\dfrac{\dot{R}(r,t)}{R(r,t)}=\dfrac{\dot{a}(r,t)}{a(r,t).}
        \label{eq:Hpp}
      \end{align}
      Obviously, at the centre of symmetry $H_\parallel(r=0,t)\equiv H_\perp(r=0,t)$. Then, the local expansion rate at the present time, $t_0$, is given by
      \begin{equation}
        H_0=H_\parallel(r=0,t_0)=H_\perp(r=0,t_0)
        \label{eq:LTB_H0}
      .\end{equation}
As in the FLRW metric, $r$ is an adimensional coordinate and not a measure of a physical distance. Being a flag coordinate, we have the freedom of rescaling $r$ such that, for example, $R_0(r)=c\, t_0\times r$. As we are considering an isotropic, but inhomogeneous fluid, all its parameters such as pressure, density, and EoS parameter $w$ may in principle depend not only on time, but also on the radial coordinate. For LTB models we consider the cosmic fluid with two components: the pressureless inhomogeneous cold matter (with $w\equiv0$) and the DE fluid (with trivial EoS parameter $w\equiv-1$). Such a DE fluid component is equivalent to having an inhomogeneous $\Lambda$ term that is constant in time and on every sphere of radius $r$, but with a possible radial profile $\Lambda(r)$. Recalling the Birkhoff theorem, we can expect that each shell evolves independently of the others, as a FLRW model with the same values of the fluid parameters. Hence, solving EE leads to the analogue of the Friedmann equation 
      \begin{equation}
        {H^2_\perp(r,t)}=H_0^2(r)\left[\dfrac{\Omega_m(r)}{a(r,t)^3}+\dfrac{\Omega_k(r)}{a(r,t)^2}+\Omega_\Lambda(r)\right]
        \label{eq:Feq}
      .\end{equation}
      Here $H_0(r)\equiv H_\perp(r, t_0)$. As in the FLRW model, $\Omega_m(r)\equiv8\pi G\rho_m(r)/3H_0^2(r)$, $\Omega_\Lambda(r)\equiv\Lambda(r) c^2/3H_0^2(r)$ and $\Omega_k(r)\equiv-k(r) c^2/[H_0^2(r) R_0^2(r)]$ are rescaled densities in units of the critical density. We still
have for each shell $\Omega_m(r) +\Omega_k(r)+\Omega_\Lambda(r) \equiv 1$. \Cref{eq:Feq} is a differential equation for $a(r,t)$ [cf. \Cref{eq:Hpp}]. In the most general case, it can be solved only numerically. Here we concentrate our analysis on two special cases for which \Cref{eq:Feq} can be solved analytically, namely $\Omega_\Lambda(r)\equiv0$ and $\Omega_k(r)\equiv0$. The solution of the former is
      \begin{align}
        &\left.
        \begin{aligned}
          a(r,t)&=\dfrac{\Omega_m(r)}{2 \Omega_k(r)}(\cosh \eta-1) \\
          t-t_B(r)&=\dfrac{1}{2H_0(r)}\dfrac{\Omega_m(r)}{\Omega_k^{3/2}(r)}(\sinh \eta-\eta)
          \label{eq:solK<0}
        \end{aligned}
        \,\right\}&&\Omega_k(r)>0\\
        &\left.
        \begin{aligned}
          a(r,t)&=\dfrac{\Omega_m(r)}{2 |\Omega_k(r)|}(1-\cos u)\\
          t-t_B(r)&=\dfrac{1}{2H_0(r)}\dfrac{\Omega_m(r)}{|\Omega_k(r)|^{3/2}}(u-\sin u)\phantom{h}
          \label{eq:solK>0}
        \end{aligned}
        \,\right\}&&\Omega_k(r)<0\\
        &\hspace{1.42em}a(r,t)=\left(\dfrac{3}{2}H_0(r)(t-t_B(r))\right)^{2/3}
        &&\Omega_k(r)=0,
        \label{eq:solK=0}
      \end{align}
      where $\eta=\eta(r,t)$ and $u=u(r,t)$ are dimensionless parameters related to the conformal time, while $t_B(r)$ is the Big Bang time of a given shell. We chose a homogeneous age of the Universe (in particular, $t_B(r)\equiv0$), as it has been shown that a Big Bang time different from shell to shell introduces decaying modes \citep{Zibin08}, which in turn implies large CMB spectral distortions \citep{Zibin11}. In the second simple case of the flat Universe $\Omega_k(r)\equiv0,$ we find a solution      \begin{equation}
        a(r,t)=\sqrt[3]{\frac{\Omega_m(r)}{\Omega_{\Lambda}(r)} \sinh^2\left[\frac{3}{2}H_0(r)t\sqrt{\Omega_{\Lambda}(r)}\right]}
        \label{eq:solDE} 
      .\end{equation}
      Obviously, as $\Omega_k(r)\equiv0$, the profile of the $\Lambda$ term is defined by the profile of matter density. This case recovers the $\Lambda$CDM model for $\Omega_m(r)\equiv {\rm const}$. We here consider LTB models with a specific matter density profile:
      \begin{equation}
       \Omega_m(r)=\Omega_{\rm out}-(\Omega_{\rm out}-\Omega_{\rm in})e^{-r^2/2\rho^2},
        \label{eq:Gaussian}
      \end{equation}
        where $\Omega_{\rm in}\leq\Omega_{\rm out}$ are the density parameters at the centre and beyond this underdensity (also called void\textup{}), while the parameter $\rho$ defines its size. Clearly, for $r\to\infty,$ we recover a FLRW model with $\Omega_m=\Omega_{\rm out}$. We fixed the value of $\Omega_{\rm out}$ to unity for consistency with the inflationary paradigm. This simple density profile provides a smooth transition from the local to the distant matter density, without introducing too many free parameters. Furthermore, we assumed the observer to be located at the centre of the void. This is obviously a privileged position, against the Copernican principle, but the assumption can be relaxed with some complication of the mathematical formalism \citep{Alnes06b}. We will address these aspects in a forthcoming paper.
        
        The two physical observables of interest, radial cosmic expansion rate and luminosity distance, are derived from observing photons that arrive radially in the chosen reference frame. Therefore, we used the relation of the two independent coordinates $r$ and $t$ with the redshift $z$ along the radial null geodesic \citep{Enqvist07}
      \begin{align}
        \dfrac{dr}{dz}&=\dfrac{c\sqrt{1-k(r)}}{(1+z)\dot{R}'(r,t)} \label{eq:drdz} \\
        \dfrac{dt}{dz}&=-\dfrac{R'(r,t)}{(1+z)\dot{R}'(r,t).} \label{eq:dtdz}
      \end{align}
      Finally, the solutions of \Cref{eq:drdz,eq:dtdz} can be used in combination with \Cref{eq:Hpl,eq:Hpp} to relate any observable as a function of the redshift. The luminosity and the angular diameter distances in LTB model (see \citet{Ellis07}) are given by
      \begin{equation}
        d_A(z)=R(r(z),t(z)),\hspace{1em}d_L(z)=(1+z)^2d_A(z)
        \label{eq:17}
      \end{equation}
      and can be calculated numerically as functions of the redshift using the equations above.
      
\section{Data analysis: methods}
    \label{sec:analysis}
    We tested the theoretical models described in \Cref{sec:th} against a number of independent observables that are SN Ia, OHD, and BAO. Our goal is to address specifically the $H_0$ determination out of these measurements, also discussing the estimates of the other cosmological parameters. We analysed each of the datasets separately and then jointly to improve the sensitivity of our estimates.
    
    \subsection{SN Ia}
    The first dataset we consider is the JLA sample of SN Ia, for the reason  we gave in the introduction. After a SN Ia is identified, the raw measurements are corrected for the Galactic extinction and the cross-filter change from observed band to the SN rest-frame B band. The JLA catalogue is built by processing these data with the use of the SALT light curve and spectral fitters \citep{Guy07}, which provide the data points that can further be used in a cosmological study. Each SN is characterised by its redshift $z$, the value for the maximum B-band apparent magnitude $m_B$, then stretch and colour correction factors, $s$ and $c$, respectively. Although considered the best-known high-redshift standard candles in cosmology, SN Ia still show small variations in their maximum absolute luminosity, and hence, in the B-band absolute magnitude, $M_B$. To take this into account, the maximum absolute magnitude has been so far corrected through the empirical relation 
  \begin{equation}
    M_B^{\rm corr}=M_B-\alpha s+\beta c,
    \label{eq:Phillips}
  \end{equation}
where $M_B^{\rm corr}$ and the two correction parameters, $\alpha$ and $\beta$, are  assumed to be constant for all the SN Ia \citep{Hamuy95, Kasen07}. The distance modulus $\mu\equiv m_B-M_B$ is related to the luminosity distance, $d_L=d_H D_L$, which contains all the cosmological information. Here, $d_H\equiv c/H_0\simeq3000 h^{-1}$Mpc is the Hubble radius, while $D_L$ is a ``Hubble constant-free'' dimensionless luminosity distance, which depends on the other cosmological parameters. Therefore, the $\mu-d_L$ relation can be written as follows
  \begin{equation}
    m_B-M_B^{\rm corr}+\alpha s-\beta c=5\log_{10}d_H + 5\log_{10}D_L +25
    \label{eq:dist-magn}
  .\end{equation}
 This equation has so far been used to simultaneously fit the three nuisance parameters ($\alpha$, $\beta$ and ${\cal M}\equiv M_B^{\rm corr}+5\log_{10} d_H$) together with the cosmological parameters relevant for the calculation of $D_L(z)$. It is evident that the distance moduli provided in SN datasets, depending on such estimates of the nuisance parameters, are consequently biased as a result of the pre-assumption of the $\Lambda$CDM model. It follows from the definition of $\cal M$ that the estimates of the cosmological parameters are insensitive to the actual value of $H_0$, which plays the role of an overall offset in \Cref{eq:dist-magn}.
 
Assuming $\alpha$, $\beta,$ and $M_B^{\rm corr}$ as constant means that every SN has the same corrected absolute magnitude and that no further corrections are needed. This statement has been questioned by \citet{Nielsen15}, who extended the procedure to consider the variation of the corrected absolute magnitude $M_B^{\rm corr}$ from one to another SN. For the sake of simplicity, $M_B^{\rm corr}$, $s,$ and $c$ are assumed to be independent Gaussian variables with normal distributions ${\cal N}(M_0,\sigma_{M_0})$, ${\cal N}(s_0,\sigma_{s_0})$ and ${\cal N}(c_0,\sigma_{c_0})$, respectively. In contrast, $\alpha$ and $\beta$ are still considered constant coefficients of \Cref{eq:Phillips}. Here we follow the same approach.
    Therefore, the joint probability for the values of the intrinsic SN parameters can be written as    
    \begin{equation}
      p(Y)=|2\pi \Sigma_l |^{-1/2} \exp[-(Y-Y_0)\Sigma_l^{-1}(Y-Y_0)^{\rm T}/2],
    \end{equation}
    where $Y=(M_1, s_1,c_1,\hdots,M_N,s_N,c_N)$ is a $3N$-vector of the true values of the intrinsic parameters for each of the $N$ supernovae; $Y_{0}=(M_0, s_0,c_0,\hdots, M_0, s_0, c_0)$ is a $3N$-vector with the central values of the parameter distributions; $\Sigma_l={\rm diag}(\sigma^2_{M_0},\sigma^2_{s_0},\sigma^2_{c_0},\hdots,\sigma^2_{M_0},\sigma^2_{s_0},\sigma^2_{c_0})$ is the $3N\times3N$ covariance matrix. 
  The measured values ($\hat m_B$, $\hat s$, $\hat c$) are conveniently expressed with the $3N$-vector $\hat Z=(\hat m_{B1}-\mu_1, \hat s_1, \hat c_1,\hdots)$. Following \citet{Nielsen15}, we define as $\theta$ the full set of free parameters, which describe both the astrophysical properties of SN and the cosmology. Then, the likelihood of the observed values, given the theoretical model $\theta$, can conveniently be written as
    \begin{align}
      {\cal L}_{\rm SN}(\hat Z|\theta)&=|2\pi(\Sigma_d+A^{\rm T} \Sigma_l A)|^{-1/2}\times\nonumber\\
      &\hspace{-1em}\exp[-(\hat Z-Y_0 A)(\Sigma_d+A^{\rm T}\Sigma_l A)^{-1}(\hat Z-Y_0 A)^{\rm T}/2],
    \end{align}
     where $\Sigma_d$ is the covariance matrix of the data \citep{Betoule14}, and $A$ is the $3N\times3N$ block-diagonal matrix
    \begin{equation}
      A=\begin{pmatrix}
    1 & 0 & 0 &   &  \\
    -\alpha & 1 & 0 &   &  \\
    \beta & 0 & 1 &   &  \\
     &  &  & \ddots &
        \end{pmatrix}.
    \end{equation}
In analysing the JLA dataset by itself, we did not assume any prior on $H_0$ because we estimated the total offset $\cal M$. The distribution ${\cal N}(M_0,\sigma_{M_0})$ implies the distribution of ${\cal M}$: ${\cal N}({\cal M}_0,\sigma_{{\cal M}_0})$, with $\sigma_{{\cal M}_0}=\sigma_{M_0}$ and ${\cal M}_0=M_0+5\log d_H$. Hence, using this method, eight parameters describe the physics of SN Ia: $\alpha$, $\beta$, ${\cal M}_0$, $\sigma_{{\cal M}_0}$, $s_0$, $\sigma_{s_0}$, $c_0$, and $\sigma_{c_0}$ (instead of three parameters, as in the conventional approach), which need to be fitted simultaneously with the cosmological parameters.
        \subsection{OHD}
    We have shown in \Cref{sec:th}. that the radial cosmic expansion rate is proportional to $dz/dt$ for FLRW (cf. \Cref{eq:defH}) and LTB models (cf. \Cref{eq:Hpl,eq:dtdz}). Therefore, by measuring the age difference of two objects at two close redshift points, we can estimate the radial expansion rate at the corresponding redshift. For this purpose, \citet{Jimenez02} proposed what is called the differential age (DA) method, which uses pairs of passively evolving red galaxies found at very similar redshifts. Another way to determine the expansion history of the Universe is using BAO \citep{Gaztanaga09, Delubac15,Sahni14}. We here
used the 23 DA points from the dataset collected and updated by \citet{Ding15}. The shape of the $H(z)$ curve constrains the cosmological models, while the offset value provides the local cosmic expansion rate, $H_0$. For this dataset, we performed a simple likelihood analysis considering that all the data points are uncorrelated, 
\begin{equation}
{\cal L}_{\rm OHD}\propto\exp\left[-\dfrac{1}{2}\sum_{i=1}^{23} \left(\frac{\hat H_i-H(z_i)}{\sigma_{H_i}}\right)^2\right]
.\end{equation}
Here $\hat H_i$ is the observed value at redshift $z_i$ with its own uncertainty $\sigma_{H_i}$. For the theoretical prediction we used for $H(z_i)$ either \Cref{eq:Hofz} for the FLRW models or \Cref{eq:Hpl} for LTB cosmologies.
\subsection{BAO}
   The large-scale structure of the Universe has been extensively studied through redshift surveys, such as the six-degree field galaxy survey 6dFGS \citep{Beutler11} and the Sloan Digital Sky Survey-SDSS \citep{Marin15}. The estimated galaxy correlation function shows a peak at large scales that is interpreted as the signature of the baryon acoustic oscillation in the relativistic plasma of the early Universe.  In FLRW cosmology, the acoustic-scale distance ratio is defined by $\Xi\equiv r_d/D_V(z)$, where 
  \begin{equation}
    r_d =\int_{z_d}^{\infty} \frac{c_s(z)}{H(z)}dz
    \label{eq:r_d}
  \end{equation}
is the comoving sound horizon at the redshift $z_d$ of the baryon drag epoch \citep{Eisenstein98}, and 
\begin{equation}
D_V(z)\equiv \left[(1+z)^2 d_A^2(z)\dfrac{cz}{H(z)}\right]^{1/3}
\label{eq:D_v}
\end{equation}
is a spherically averaged distance measure introduced by \citet{Eisenstein05}. In LTB cosmology, structure formation is poorly understood, which
is the reason why using the BAO data for this model is still controversial \citep{Zibin08, Clarkson09}.

We used the measurements from the 6dFGS \citep{Beutler11}, the SDSS DR7 \citep{Ross15}, and the BOSS DR11 \citep{Anderson14} samples and the Ly$\alpha$ forest measurements from BOSS DR11 \citep{Delubac15,Ribera14}. The 6dFGS sample we used contains 75,117 galaxies up to  $z<0.15$ and gives $r_d/D_V=0.336\pm0.015$ at an effective redshift $z_{eff}=0.106$. The SDSS DR7 catalogue contains 63,163 galaxies at $z < 0.2$ and the measurement given is $D_V(z_{eff}=0.15)=(664\pm25)(r_d/r_{d,fid})$ Mpc, where $r_{d,fid}=148.69$ Mpc in their fiducial cosmology. BOSS DR11 contains nearly one million galaxies in the redshift range $0.2 < z < 0.7$ and the two measurements given are $D_V(z_{eff}=0.32)=(1264\pm25)(r_d/r_{d,fid})$ Mpc and $D_V(z_{eff}=0.57)=(2056\pm20)(r_d/r_{d,fid})$ Mpc with $r_{d,fid}=149.29$ Mpc. The Ly$\alpha$ forest of BOSS DR11 consists of 137,562 quasars in the redshift range $2.1\leq z \leq 3.5$ and provides $d_A/r_d=11.28\pm0.65$ , $d_H/r_d=9.18\pm0.28$ at $z_{eff}=2.34$, and $d_A/r_d=10.8\pm0.4$ , $d_H/r_d=9.0\pm0.3$ at $z_{eff}=2.36$, where $d_H(z)=c/H(z)$. To homogenise the measured BAO quantity as $r_d/D_V(z)$, we inverted the results of SDSS DR7 and BOSS DR11 using their fiducial values $r_{d,fid}$ and combined the measurements of Ly$\alpha$ forest from BOSS DR11 by means of  \Cref{eq:D_v}. The observed values $\hat\Xi(z)\equiv r_d/D_V(z)$  are shown in \Cref{tab:BAO}.

\begin{table}[h]
\begin{center}
\footnotesize 
\caption{BAO data.}
\begin{tabular}{|c|c|c|c|}
\hline 
 $z$ &$\hat \Xi$ & $\sigma_\Xi$ & Ref.  \\
\hline
 0.106 & 0.336 &  0.015    &  \citet{Beutler11}\\
\hline
0.15 & 0.2239 & 0.0084  & \citet{Ross15} \\
\hline
0.32 &  0.1181 & 0.0023 & \citet{Anderson14} \\
\hline
0.57 & 0.0726 & 0.0007  & \citet{Anderson14} \\
\hline
2.34 & 0.0320 & 0.0016  & \citet{Delubac15} \\
\hline
2.36 &  0.0329 & 0.0012 & \citet{Ribera14}\\
\hline
\end{tabular}
 \label{tab:BAO}
\end{center}
\end{table}

The acoustic scale has also been measured by \citet{Kazin14} using the WiggleZ galaxy survey. They reported three correlated measurements at redshifts 0.44, 0.60, and 0.73. However, we decided not to use these measurements because the WiggleZ volume partially overlaps that of the BOSS sample and the correlations between the two surveys have not been quantified.

As all the BAO data points in \Cref{tab:BAO} are uncorrelated, we performed a simple likelihood analysis:
\begin{equation}
 {\cal L}_{\rm BAO}\propto\exp\left[-\dfrac{1}{2} \sum_{i=1}^{6} \left(\frac{\hat \Xi_i-\Xi(z_i)}{\sigma_{\Xi_i}}\right)^2\right]
.\end{equation}
Here $\hat \Xi_i$ is the observed value at redshift $z_i$ with its own uncertainty $\sigma_{\Xi_i}$ (see \Cref{tab:BAO}). The theoretical estimate, $\Xi(z_i)$, has been evaluated using \Cref{eq:D_v,eq:r_d} and by exploiting the fitting formula for $z_d$ given in \citet{Eisenstein98}.
       \subsection{Combined analysis}
We show in the next section that each of the three datasets we considered provides good estimates of the cosmological parameters. However, it is worthwhile combining all three datasets to obtain more stringent constraints. Moreover, the joint analysis allows us to provide separate estimates for the SN absolute magnitude and the Hubble constant. Assuming that the three datasets are independent, we evaluated the total likelihood as the product of the likelihoods of the single datasets. Therefore, for FLRW models $\cal{L}_{\rm tot}=\cal{L}_{\rm SN}\cal{L}_{\rm OHD}\cal{L}_{\rm BAO}$, while for LTB cosmologies $\cal{L}_{\rm tot}=\cal{L}_{\rm SN}\cal{L}_{\rm OHD}$ because in this case we did not consider the BAO dataset.
  \section{Data analysis: results}
    \label{sec:res}
      We present the results for cosmological models based on the FLRW and the LTB metrics. Several FLRW models have been analysed by \citet{Nielsen15} using the JLA dataset, and we fully agree with their results. We extend their discussion to the $w$CDM and LTB models, and also present the constraints derived from the OHD and BAO datasets. The JLA, OHD, and BAO data strongly constrain the apparent acceleration of the cosmic expansion, the current expansion rate, and the curvature of the Universe, respectively. A joint analysis, then, allows us to see what improvement we may have in the estimation of the cosmological parameters.
   \subsection{Constraints on flat $\Lambda$CDM model}
    The flat $\Lambda$CDM model has so far been tested with the available cosmological observables and is commonly considered the concordance model in cosmology.
      \begin{table}[h!]
        \begin{center}
          \footnotesize
          \caption{Results from the fits of the $\Lambda$CDM model to the data.}
          \begin{tabular}{|@{}c@{}|@{\phantom{..}}c@{\phantom{..}}|c|c@{}|}
            \Xhline{2\arrayrulewidth}
            Data & $H_0\,${\fontsize{8.2}{0}\selectfont[${\rm km}/{\rm s}/$Mpc]}& $\Omega_m$ & $\Omega_{\Lambda}=1-\Omega_m$\\
            \Xhline{2\arrayrulewidth}
            {\fontsize{8.5}{0}\selectfont JLA} & - & $0.376\pm 0.031$ & $0.624 \pm 0.031\,\,\,$\\
            \hline
            {\fontsize{8.5}{0}\selectfont OHD}&$68.7\pm3.3$&$ 0.319\pm 0.061$&$0.681\pm0.061\,\,\,$\\
            \hline
            {\fontsize{8.5}{0}\selectfont BAO}&$67.3\pm2.2$&$0.334\pm 0.042$&$0.689\pm0.037\,\,\,$\\
            \hline
            {\fontsize{8.5}{0}\selectfont JLA+OHD}&$66.7\pm2.0$&$0.366\pm0.028$&$0.634\pm0.028\,\,\,$\\
            \hline
            {\fontsize{8.5}{0}\selectfont $\,\,$JLA+OHD+BAO$\,\,$}&$67.8\pm1.0$&$0.350\pm0.016$&$0.650\pm0.016\,\,\,$\\
            \hline
          \end{tabular}
          \label{tab:LCDM}
        \end{center}
      \end{table}
      The results of our analysis for this model are presented in \Cref{tab:LCDM}. In particular, using the JLA dataset alone, we reproduce the $\Omega_m$ value found by \citet{Nielsen15}. In addition to their results, we also quote the uncertainties in the parameter estimates, which show that the JLA dataset provides consistent results with those we find for OHD and BAO samples. However, our values for $\Omega_m$, obtained from JLA alone and JLA+OHD+BAO are both different by about 1.9$\sigma$  from the most recent determination of Planck (TT, TE, EE + lowP), which provided $\Omega_m=0.316\pm0.009$.
      \begin{figure}
        \stepcounter{figure}
        \includegraphics[width=\columnwidth]{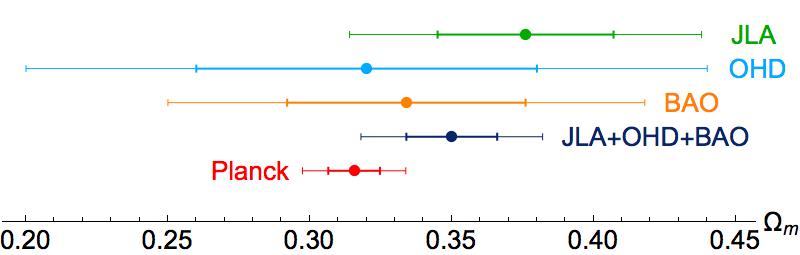}\par
        \Scaption{Constraints on matter density}
        \vspace{1em}
        \includegraphics[width=\columnwidth]{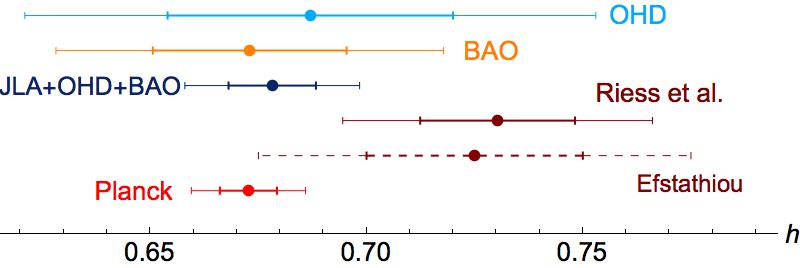}
        \Scaption{Constraints on Hubble constant}
        \vspace{.5em}
        \addtocounter{figure}{-1}
        \caption{Results at the $1\sigma$ and $2\sigma$ c.l. for the parameters of the $\Lambda$CDM model when fitted to JLA (green), OHD (azure), BAO (orange), and the three datasets combined (dark blue). Constraints from the direct measurement by \citet{Riess16} (dark red), the reanalysis by \citet{Efstathiou14} (dashed dark red), and the \citet{Planck15} (red) are also shown.}
        \label{fig:LCDM}
      \end{figure}
      
      The $H_0$ estimates derived from the OHD and BAO data for the current expansion rate
are consistent by themselves and with Planck.  The JLA dataset by itself is insensitive to  $H_0$. However, in the joint analysis JLA  constrains other parameters of the fit, which in turn affect the estimate of $H_0$. The final value for $H_0$ derived from the joint JLA+OHD+BAO analysis is in excellent agreement with the value from Planck (\Cref{fig:LCDM}), but is still different
by $1.7\sigma$ and $2.5\sigma$  from the findings by \citet{Efstathiou14} and \citet{Riess16}. 
    
    \subsection{Constraints on $k\Lambda$CDM model}
\begin{figure}[h!]
        \includegraphics[width=\columnwidth]{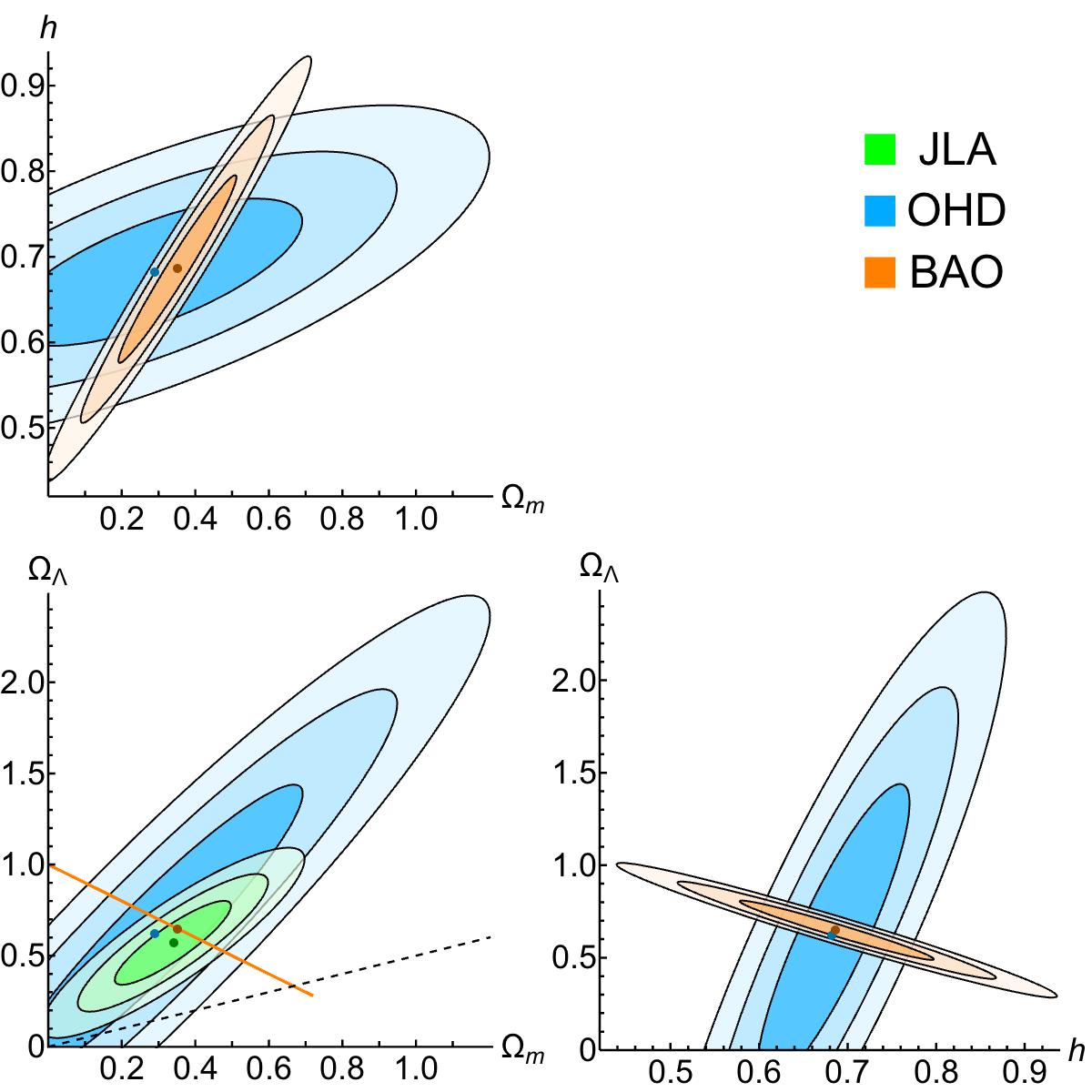}
        \caption{$1\sigma$, $2\sigma$ and $3\sigma$ confidence regions resulting from the fit of $k\Lambda$CDM model to the single datasets as indicated in the top right panel. The dashed line in the $\Omega_m-\Omega_\Lambda$ plane represents the transition from the decelerating (below) to the accelerating (above) models.}
        \label{fig:FLRW_ell}
      \end{figure}

      This model is completely defined by the three cosmological parameters $H_0$, $\Omega_m$ , and $\Omega_\Lambda$. The best-fit values together with their statistical errors in \Cref{tab:FLRW} show a clear consistency of the results obtained by analysing the single datasets.
      \begin{table}[h!]
        \begin{center}
          \footnotesize
          \caption{Results from the fits of the $k\Lambda$CDM model to the data.}
          \begin{tabular}{|@{}c@{}|@{\phantom{..}}c@{\phantom{..}}|c|c@{}|}
            \Xhline{2\arrayrulewidth}
           Data & $H_0\,${\fontsize{8.2}{0}\selectfont[${\rm km}/{\rm s}/$Mpc]}& $\Omega_m$ & $\Omega_{\Lambda}$\\
            \Xhline{2\arrayrulewidth}
            {\fontsize{8.5}{0}\selectfont JLA} & - & $0.341\pm 0.098$ & $0.569 \pm 0.149\,\,\,$ \\
            \hline
            {\fontsize{8.5}{0}\selectfont OHD}&$68.2\pm5.7$&$0.291 \pm 0.265$&$0.622 \pm 0.539\,\,\,$\\
            \hline
            {\fontsize{8.5}{0}\selectfont BAO}&$68.7 \pm 7.3$&$0.354\pm0.106$&$0.646\pm0.106\,\,\,$\\
            \hline
            {\fontsize{8.5}{0}\selectfont JLA+OHD}&$66.3 \pm 2.2$&$0.319 \pm 0.044$&$0.556 \pm 0.144\,\,\,$\\
            \hline
            {\fontsize{8.5}{0}\selectfont $\,\,$JLA+OHD+BAO$\,\,$}&$68.1\pm1.0$&$0.350\pm0.016$&$0.650\pm0.016\,\,\,$\\
            \hline
          \end{tabular}
          \label{tab:FLRW}
        \end{center}
      \end{table}
We present the confidence regions for the $k\Lambda$CDM model in \Cref{fig:FLRW_ell}. Our results show that the BAO data constrain the curvature of the Universe quite strictly to be zero, leading to a strong correlation between $\Omega_m$ and $\Omega_\Lambda$ and shrinking the confidence regions to a line in $\Omega_m-\Omega_\Lambda$ plane. An alternative way to study the BAO data is by parametrising the $k\Lambda$CDM model with $\Omega_m$ and $\Omega_k$. In this case, the result for $\Omega_m$ is the same as in \Cref{tab:FLRW}, while the best-fit value of $\Omega_k$ is vanishingly small and consistent with zero. For this reason, the best-fit values of $H_0$ and $\Omega_m$ are not exactly the same as those obtained for $\Lambda$CDM (cf. \Cref{tab:LCDM}), although they agree very well. The constraints from the JLA and OHD samples are fully consistent with the flat cosmology, although the best-fit values slightly differ from this line. The confidence regions in the $\Omega_m-h$ and $h-\Omega_\Lambda$ planes, resulting from the OHD and BAO data, fully overlap. The estimates for $H_0$ derived from the BAO and OHD analysis are consistent by themselves (see \Cref{tab:FLRW}). In our full joint analysis, the $k\Lambda$CDM and the $\Lambda$CDM models provide almost the same parameter estimates.
      \subsection{Constraints on $w$CDM model}
\begin{figure}[h!]
        \includegraphics[width=\columnwidth]{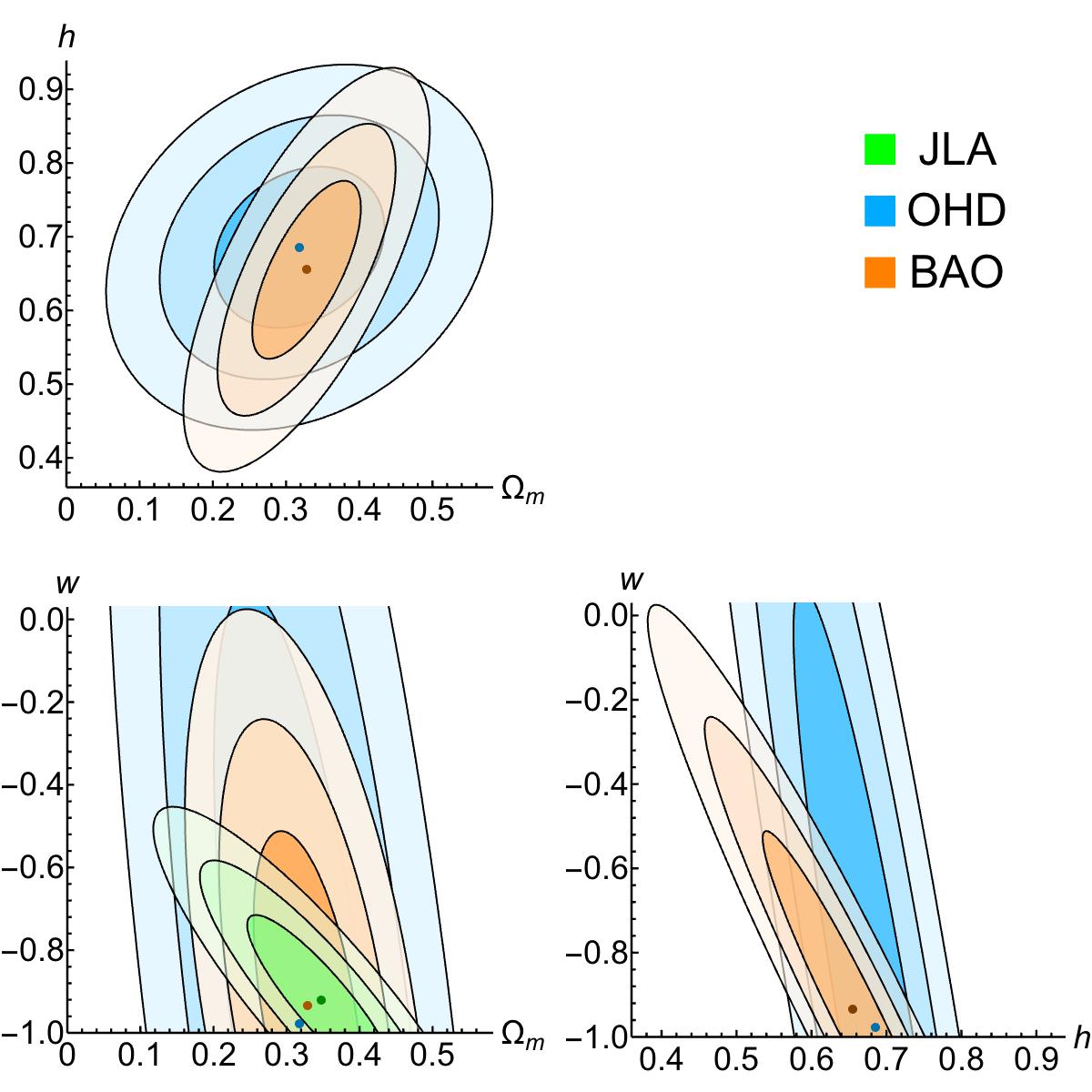}
        \caption{$1\sigma$, $2\sigma$ and $3\sigma$ confidence regions from the fits of the $w$CDM model to the single datasets as indicated in the top right panel.}
        \label{fig:wCDM_ell}
      \end{figure}

The $w$CDM model considers a DE fluid with a free EoS parameter $w$ instead of a constant $\Lambda$ term corresponding to $w=-1$. The addition of one more parameter implies larger error bars in the $\Omega_m$ and $H_0$ determinations presented in \Cref{tab:wCDM}.
      \begin{table}[h!]
        \begin{center}
          \footnotesize 
          \caption{Results from the fits of the $w$CDM model to the data.}
          \begin{tabular}{|@{}c@{}|@{\phantom{.}}c@{\phantom{.}}|c|@{\phantom{.}}c@{\phantom{.}}|}
            \Xhline{2\arrayrulewidth} 
            Data & $H_0\,${\fontsize{8.5}{0}\selectfont[${\rm km}/{\rm s}/$Mpc]}& $\Omega_m$ & $w$\\
            \Xhline{2\arrayrulewidth}
            JLA & - & $0.347\pm 0.119$ & $-0.92 \pm 0.30$ \\
            \hline
            OHD&$68.5\pm7.2$&$0.318\pm0.077$&$-0.98\pm0.69$\\
            \hline
            BAO&$65.5\pm8.0$&$0.329\pm0.049$&$-0.93\pm0.28$\\
            \hline
            JLA+OHD&$67.0\pm1.9$&$0.318\pm0.073$&$-0.86\pm0.17$\\
            \hline 
            $\,\,$JLA+OHD+BAO$\,\,$&$66.5\pm1.8$&$0.346\pm0.017$&$-0.93\pm0.07$\\
            \hline 
          \end{tabular}
          \label{tab:wCDM}
        \end{center}
      \end{table}
      The confidence regions derived from the single datasets are consistent among themselves (see \Cref{fig:wCDM_ell}). In particular, the estimates of $w$ are consistent with $w=-1$, the $\Lambda$CDM model. The value of $\Omega_m$ resulting from the full joint analysis is very close to the value obtained for the $\Lambda$CDM model. The OHD and BAO estimates of $H_0$ are poorly constrained.  However, again, the joint JLA+OHD+BAO analysis provides a value of $H_0$ that is different by $1.7\sigma$ and $2.6\sigma$   from the results of  \citet{Efstathiou14} and \citet{Riess16} , respectively. 
    \subsection{Constraints on LTB model}
\begin{figure}[h!]
        \includegraphics[width=\columnwidth]{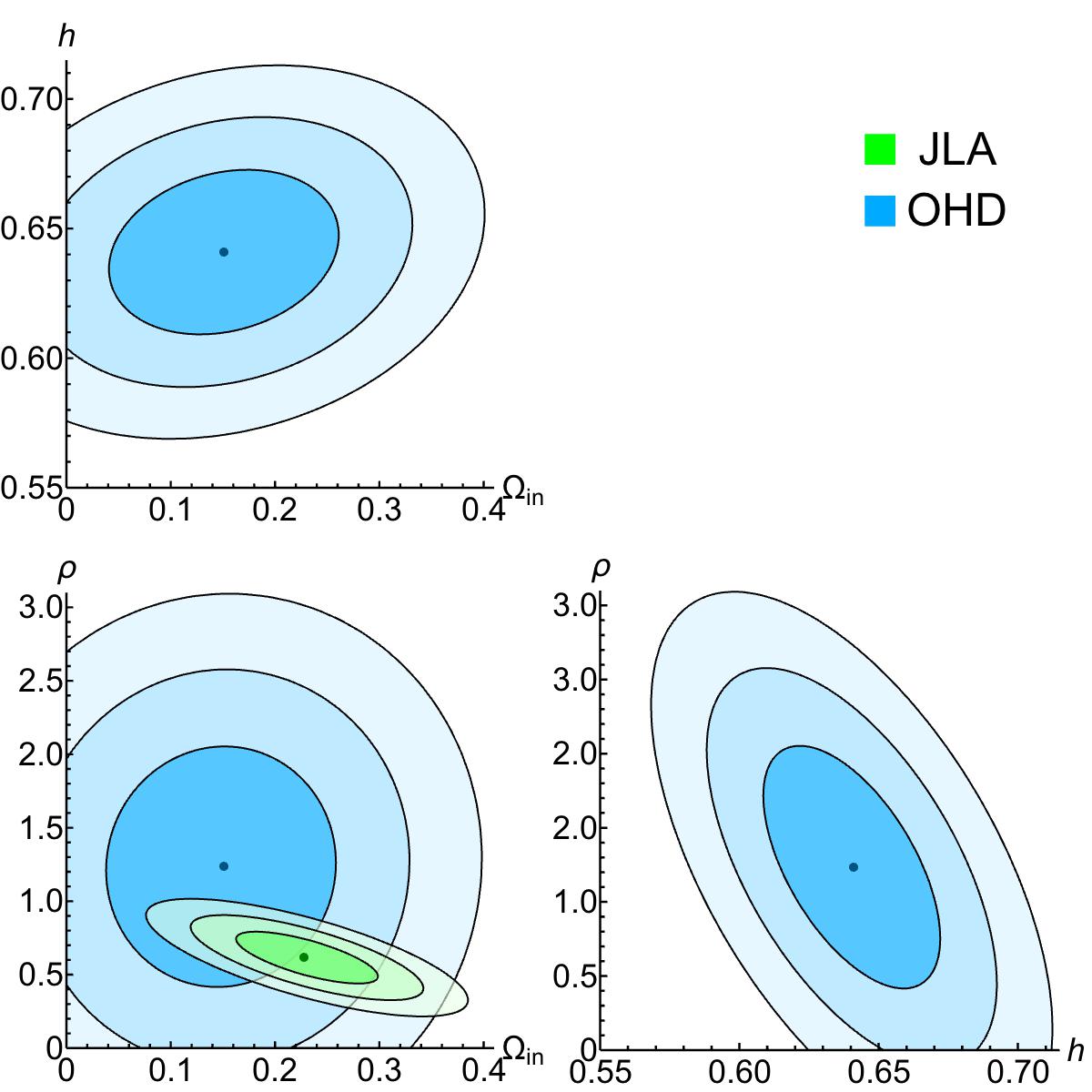}
        \caption{$1\sigma$, $2\sigma$ and $3\sigma$ confidence regions for the LTB model to the single datasets as indicated in the top right panel.}
        \label{fig:LTB_ell}
      \end{figure}

    Here we extend the discussion of \citet{Nielsen15} to consider LTB models. In the LTB model, the apparent acceleration of the local Universe arises because the matter density decreases radially from high to local redshifts. The radial matter density profile is in our case completely defined by \Cref{eq:Gaussian}, where we fixed $\Omega_{\rm out}=1,$ and the remaining free parameters are the local value of the matter density $\Omega_{\rm in}$ and the dimensionless parameter $\rho$, related to the size of the void. For the JLA data, these two are the only cosmological parameters in the fit because  $H_0$  is included in the offset $\cal M$. Instead, for the OHD data we can fit all the three cosmological parameters. 
      \begin{table}[h!]
        \begin{center}
          \footnotesize 
          \caption{Results from the fits of the LTB model to the data.}
          \begin{tabular}{|c|c|c|c|}
            \Xhline{2\arrayrulewidth}
            Data & $H_0${\fontsize{8}{0}\selectfont[${\rm km}/{\rm s}/$Mpc]}& $\Omega_{\rm in}$ & $\rho$ \\
            \Xhline{2\arrayrulewidth}
            JLA & - & $0.228\pm 0.046$ & $0.61 \pm 0.13$ \\
            \hline
            OHD&$64.1\pm3.1$&$0.151 \pm 0.073$&$1.23\pm0.54$\\
            \hline
            $\,\,$JLA+OHD$\,\,$&$64.2\pm1.9$&$0.174\pm0.038$&$0.86\pm0.19$\\
            \hline
          \end{tabular}
          \label{tab:LTB}
        \end{center}
      \end{table}
      The JLA sample constrains $\Omega_{\rm in}$ to be about a quarter of the critical density, while the value resulting from the OHD dataset is lower, but agrees within the errors (cf. \Cref{tab:LTB}). To gauge the physical size of the void, we considered the comoving angular diameter distance $R_0(r=\rho)=c\,t_0\,\rho$. The age of the Universe, $t_0$, in the LTB model is related to $H_0$ through \Cref{eq:solK<0}. Hence, we evaluated the angular diameter distance using $H_0=100\,$km$\,$s$^{-1}/{\rm Mpc}\times h$. For the $\rho$ best-fit value from the JLA analysis we obtain $R_0(r=\rho)\approx 2.5\,$Gpc$/h$. The OHD analysis prefers a much larger void, but with a profile consistent with the JLA data (see \Cref{fig:LTB_ell}).
      We similarly performed a joint JLA+OHD analysis for the LTB model. The $H_0$ best-fit value is lower than the value obtained for $\Lambda$CDM. This characteristic of LTB models has been observed in earlier works: our $H_0$ estimate is consistent within $2\sigma$ with the findings of \citet{Nadathur11} and at the $1\sigma$ level with the value from \citet{HST01}. However, our value differs by $2.6\sigma$ from the result of \citet{Efstathiou14} and by $3.4\sigma$ from the value given by \citet{Riess16}. These disagreements are even stronger than for the $\Lambda$CDM model.
      In our analysis we also considered the $\Lambda$LTB model, and found that it converges to $\Lambda$CDM when both JLA and OHD datasets are used (see \Cref{tab:LLTB}). The best-fit value of $\Omega_{\rm in}$ is equal to $\Omega_m$ obtained for $\Lambda$CDM, while the size of the void tends to infinity.
      \begin{table}[h!]
        \begin{center}
          \footnotesize 
          \caption{Results from the fits of the $\Lambda$LTB model to the data.}
          \begin{tabular}{|c|c|c|c|}
            \Xhline{2\arrayrulewidth}
            Data & $H_0${\fontsize{8}{0}\selectfont[${\rm km}/{\rm s}/$Mpc]}& $\Omega_{\rm in}$ & $\rho$ \\
            \Xhline{2\arrayrulewidth}
            JLA & - & $0.376\pm 0.031$ & $\infty$ \\
            \hline
            OHD&$68.7\pm3.3$&$0.319 \pm 0.061$&$\infty$\\
            \hline
            $\,\,$JLA+OHD$\,\,$&$66.7\pm2.0$&$0.366\pm0.028$&$\infty$\\
            \hline
          \end{tabular}
          \label{tab:LLTB}
        \end{center}
      \end{table}
      
To better compare the considered cosmologies, we used the Akaike information criterion (AIC) and Bayes factor ($K$). The Akaike estimate of minimum information \citep{Akaike74} for a given theoretical model and a given dataset is defined as
      \begin{equation}
        {\rm AIC} = -2\log{\cal L}^{max} + 2 p
      ,\end{equation}
      where $p$ is the number of independent parameters. By definition, this test gives preference to the model with the lowest AIC. In \Cref{tab:AIC} we present the differences, $\Delta$(AIC), of the AIC values between each theoretical scenario and $\Lambda$CDM. The Bayes factor similarly provides a criterion for choosing between two models by comparing their best likelihood values. The Bayes factor, $K={\cal L}_{\rm M_1}^{\rm max}/{\cal L}^{\rm max}_{\rm M_2}$ represents the odds for the model $M_1$ against the alternative model $M_2$. It is commonly considered  that odds lower than 1:10 indicate a strong evidence against $M_1$ \citep{Jeffreys85}. In reversed reading, odds greater than $10:1$ indicate a strong evidence against $M_2$.
 In \Cref{tab:AIC} we also show the Bayes factor of every model considered in this work against $\Lambda$CDM.
      \begin{table}[h!]
        \begin{center}
          \footnotesize 
          \caption{Comparison of the cosmological models by $\Delta({\rm AIC})={\rm AIC}_X-{\rm AIC}_{\Lambda CDM}$ and  $K={\cal L}^{\rm max}_{\Lambda {\rm CDM}}/{\cal L}_X^{\rm max}$, using the combined analysis JLA+OHD.}
          \begin{tabular}{|c|c|c|}
            \Xhline{2\arrayrulewidth}
            Model $X$& $\Delta$(AIC)&$K$\\
            \Xhline{2\arrayrulewidth}
            $\Lambda$CDM & 0 & 1\\
            \hline
            $k\Lambda$CDM&1.69&$1:1.17$\\
            \hline
            $w$CDM&1.40&$1:1.35$\\
            \hline
            LTB&9.41&$40:1$\\
            \hline
            $\Lambda$LTB&2&$1$\\
            \hline
          \end{tabular}
          \label{tab:AIC}
        \end{center}
      \end{table}
      The standard $\Lambda$CDM model is preferred by the Akaike criterion for fitting the JLA+OHD data. We note that the $k\Lambda$CDM and $w$CDM models both have one  parameter more than $\Lambda$CDM. Nevertheless, the latter still has a lower AIC value, since the Akaike criterion rewards the model with fewer parameters. The LTB model is strongly disfavoured over $\Lambda$CDM by both the AIC criterion and the Bayes factor. This behaviour  mostly arises from the SN data. From our fit to the OHD data alone we obtain $-2\log{\cal L}$ values of $12.91$ and $12.56$ for LTB and $\Lambda$CDM, from which we conclude that LTB  can be used to fit OHD data, as found by \citet{Wang12}. However, after using the JLA dataset alone with the \citet{Nielsen15} approach, we obtain  $-2\log{\cal L}$  values of $-209.88$ and $-214.83$ for the LTB and $\Lambda$CDM models, respectively. We therefore conclude that the LTB model is not performing as well as the concordance model in fitting the SN Ia data. This is in contrast to the previous findings in the literature \citep{Alnes06a,Garfinkle06, Blomqvist10}, and consistent with the work by \citet{Vargas15}. 

    \subsection{SN Ia intrinsic parameters}
    When performing the combined analysis, we were able to simultaneously
fit all the cosmological parameters and the eight intrinsic astrophysical parameters of SN Ia. The latter are those characterising the normal distributions ${\cal N}(M_0,\sigma_{M_0})$, ${\cal N}(s_0,\sigma_{s_0})$ and ${\cal N}(c_0,\sigma_{c_0})$, and the constant coefficients $\alpha$ and $\beta$ of \Cref{eq:Phillips}.
      \begin{table}[h!]
        \begin{center}
          \footnotesize 
          \caption{Results for the SN Ia intrinsic parameters from the combined JLA+OHD+BAO fit of the standard $\Lambda$CDM model.}
          \begin{tabular}{|c|c|c|c|}
            \hline 
            $M_0$ & $\sigma_{M_0}$ & $s_0$ & $\sigma_{s_0}$\\
            \hline
            $-19.13 \pm 0.04$ & 0.108 $\pm$ 0.005 & 0.038 $\pm$ 0.038 & 0.932 $\pm$ 0.027 \\
            \hline\hline
            $c_0$ & $\sigma_{c_0}$ &$\alpha$ & $\beta$\\
            \hline
            -0.016 $\pm$ 0.005 & 0.071 $\pm$ 0.002 & 0.134 $\pm$ 0.006 & 3.059 $\pm$ 0.087 \\
            \hline 
          \end{tabular}
          \label{tab:nuis}
        \end{center}
      \end{table}
      As an example, we present in \Cref{tab:nuis} estimates of these nuisance parameters  for the $\Lambda$CDM model as they
result from a combined fit with all the three datasets. The peak luminosity of SN does not have a constant value even after the corrections for the stretch and  colour factors:  the variation in the corrected SN Ia absolute magnitude is about $0.22$ at the $2\sigma$ level.  In addition to light-curve shape and  colour, the peak luminosity was similarly correlated to other parameters, including the host galaxy mass and the metallicity \citep{Kelly10,Hayden13}. Even these effects can be naturally taken into account in the distribution ${\cal N}(M_0,\sigma_{M_0})$, considering that any additional correlations are expected to decrease the distribution width. Interestingly enough, the best-fit results of the SN parameters seem to be independent of the cosmological model under consideration: changing the model at most introduces variations on the last significant digit in the numbers given in \Cref{tab:nuis}. The only exception is the central value of  ${\cal N}(M_0,\sigma_{M_0})$ obtained from the combined JLA+OHD fit for the LTB model: $M_0=-19.21\pm0.06$. This is due to the lower value of $H_0$ resulting from the fit to the OHD data.

    \section{Indirect $H_0$ estimates}
    \label{sec:H0}
\begin{figure}[h]
          \includegraphics[width=\columnwidth]{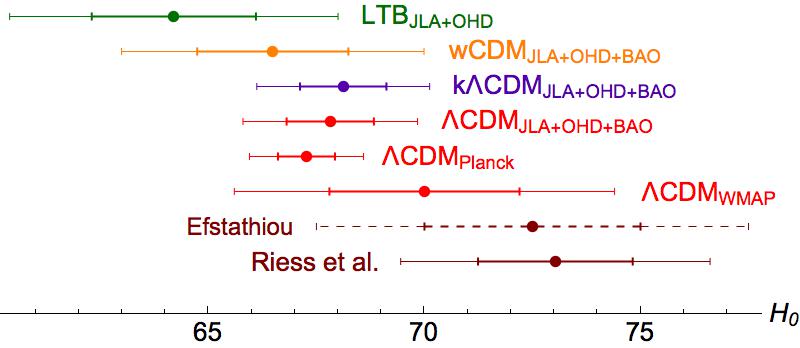}
        \caption{Results at the $1\sigma$ and $2\sigma$ c.l. for $H_0$, in standard units of km s$^{-1}/{\rm Mpc}$ from our combined analysis for LTB, $w$CDM, $k\Lambda$CDM, and $\Lambda$CDM models. The results from CMB-only measurements by the \citet{Planck15} and \citet{WMAP9} and the direct estimates by \citet{Riess16} and \citet{Efstathiou14} are also shown for comparison.}
          \label{fig:H}
        \end{figure}

The estimates of the Hubble constant from our joint analysis and from WMAP and Planck are shown in \Cref{fig:H} together with the most recent direct estimates by \citet{Efstathiou14} and \citet{Riess16}. The excellent agreement between our value of $H_0$ for $\Lambda$CDM and the CMB-only Planck measurement is remarkable, as is the consistency of the results we obtained for different FLRW cosmologies. On the other hand, the value of $H_0$ in LTB is lower than those of the FLRW models.
 
The indirect estimates of $H_0$ result to be systematically lower than the direct estimates. The best-fit values of $H_0$ for the $\Lambda$CDM and the LTB models differ by $1.7\sigma$ and $2.6\sigma$  from the value of \citet{Efstathiou14}, and differ by $2.5\sigma$ and $3.4\sigma$ from the result of \citet{Riess16}. For $\Lambda$CDM, \citet{Riess16} suggested that an additional source of dark radiation in the early Universe might allow a best-fit of the Planck data with a higher value for $H_0$. This change would certainly affect the value of the sound horizon and consequently  our $H_0$ estimate from BAO. However, this change cannot affect our result from the JLA+OHD analysis, which still differs by 2.4$\sigma$ from the results of \citet{Riess16} and completely agrees with Planck. We conclude that this difference cannot be eliminated by changing $N_{\rm eff}$ in the concordance model or by invoking possible systematic uncertainties in the CMB measurements. If this were the case, we would not have found a good agreement between our JLA+OHD+BAO result and  Planck (TT, TE, EE + lowP).
     
\begin{figure}[h!]
          \includegraphics[width=\columnwidth]{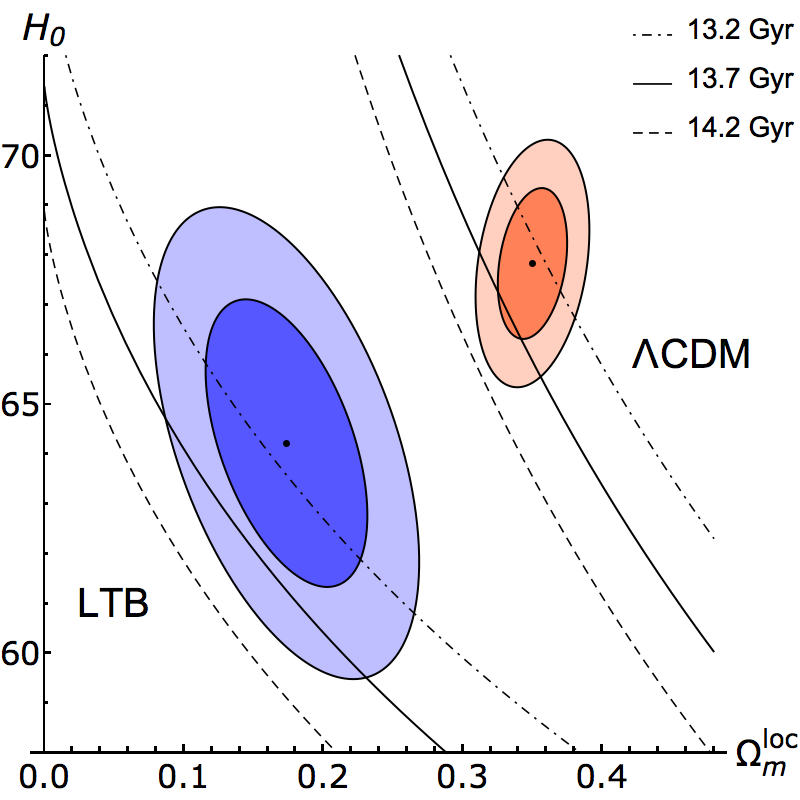}
          \caption{Theoretical iso-ages for the $\Lambda$CDM and LTB models, corresponding to ages of ($13.2\pm0.5$) Gyr are shown in the $\Omega_m^{\rm loc}-H_0$ plane, together with the 1$\sigma$ and 2$\sigma$ confidence regions resulting from the joint JLA+OHD+BAO (JLA+OHD) analysis for the $\Lambda$CDM (LTB) model.}
          \label{fig:isoch}
        \end{figure}

      In a cosmological model, the age of the Universe, $t_0$, is completely defined by the $H_0$ estimates. Here we used the result of our combined analysis for the Hubble constant to compare $t_0$ with the estimate of the absolute ages of stellar systems \citep{Bono10,Monelli15}. There is a quite good convergence on the value $t_0=(13.7\pm0.5)$ Gyr from different classes of observations (see, for a review, \citet{Freedman10}). Therefore, we show in \Cref{fig:isoch} the iso-ages corresponding to 13.2, 13.7, and 14.2 Gyr for the $\Lambda$CDM and LTB models. We show the theoretical predictions in the same $\Omega_m^{\rm loc}-H_0$ plane, where the local matter density $\Omega_m^{\rm loc}$ corresponds to $\Omega_m$ or to $\Omega_{\rm in}$ for the $\Lambda$CDM or LTB model, respectively. Our estimates for the age of the Universe derived from the best-fit results of the combined analysis described above are $t_0=(13.3\pm0.4)$ Gyr for $\Lambda$CDM and $t_0=(13.1\pm0.7)$ Gyr for LTB. In \Cref{fig:isoch} we also show the $1\sigma$ and $2\sigma$ confidence regions resulting from the JLA+OHD+BAO analysis for the $\Lambda$CDM model, and those obtained from the JLA+OHD analysis for LTB. They are completely consistent with the observational estimate of $t_0$ quoted above. 
  \section{Summary and conclusions}
    \label{sec:con}    
The combined analysis of JLA, OHD and BAO datasets allowed us to reach more stringent constraints on the cosmological parameters\footnote{Our analysis has been implemented in {\it Mathematica} 10. The code is available upon request.} and to break the degeneracy between the SN absolute magnitude and the cosmic expansion rate. Our main findings can be summarised as follows. 

By fitting the cosmological and the SN intrinsic parameters to the combined set of JLA, OHD, and BAO data, we constrained the distributions of SN absolute magnitude, stretch, and colour. The resulting values are  cosmological-model independent, with the exception of the $M_0$ value obtained for LTB.  The method we used can in principle be extended to include the effects related to mass and metallicity of the host galaxies.
 
We studied the $\Lambda$CDM model and its extensions to consider non-vanishing spatial curvature and different assumptions for the DE component. The combined analysis clearly prefers the concordance model, as it forces the curvature to vanish and the DE EoS to be consistent with $w=-1$.    

We also studied an LTB model with a Gaussian profile, which is strongly disfavoured with respect to the concordance model by information criteria, such as AIC analysis or Bayes factor.

For the $\Lambda$CDM model, the JLA+OHD+BAO analysis provides a value of $H_0=(67.8 \pm 1.0) \,$km$\,$s$^{-1}/$Mpc that is fully consistent with the Planck (TT, TE, EE + lowP) result. This means that the difference with the direct measurements by \citet{Riess16} is very likely not due to systematics in the Planck CMB measurements.  It also seems difficult to reconcile direct and indirect $H_0$ measurements by considering an additional source of dark radiation in the early Universe, as this would not affect the JLA+OHD fit, which is still consistent with Planck. Therefore, it is still unclear wether it is necessary to extend the concordance $\Lambda$CDM model.
    \begin{acknowledgements}
      We are grateful to Jeppe Tr{\o}st Nielsen for useful discussions on the JLA data analysis, and to Martin White for his advice about the BAO data. We thank Giuseppe Bono and the referee for their constructive comments and helpful suggestions. 
    \end{acknowledgements}

\bibliographystyle{aa}
\bibliography{references}

\end{document}